\documentclass{aa}

\usepackage{graphicx}
\usepackage[varg]{txfonts}
\bibpunct{(}{)}{;}{a}{}{,} % to follow the A&A style
\begin{document}

\title{Evolution of lithium abundance in the Sun and solar twins}

   \author{
      F. Th\'evenin
      \inst{\ref{inst1}}
\and
      A.V. Oreshina\thanks{visiting astronomer at OCA, France}
      \inst{\ref{inst2}}
\and
      V.A. Baturin\thanks{visiting astronomer at OCA, France}
      \inst{\ref{inst2}}
\and
      A.B. Gorshkov
      \inst{\ref{inst2}}
\and
      P. Morel
      \inst{\ref{inst1}}
\and
      J.~Provost
       \inst{\ref{inst1}}
          }

   \institute{Universit\'e de La Cote d'Azur, OCA, Laboratoire
Lagrange CNRS, BP. 4229, 06304, Nice Cedex, France\label{inst1}
 \and
Sternberg Astronomical Institute, Lomonosov Moscow State
University, Moscow, Russia\label{inst2}}

%             \email{anna.v.oreshina@gmail.com}
%             \thanks{}

   \date{Received ~~~ July, 2016; accepted }

\abstract
{Evolution of the \element[][7]{Li} abundance in the convection
zone of the Sun during different stages of its life time is
considered to explain its low photospheric value in comparison
with that of the solar system meteorites. Lithium is intensively
and transiently burned in the early stages of evolution (pre-main
sequence, pMS) when the radiative core arises, and then the
\element[][]{Li} abundance only slowly decreases during the main
sequence (MS). We study the rates of lithium burning during these
two stages.

In a model of the Sun, computed ignoring pMS and without
extra-convective mixing (overshooting) at the base of the
convection zone, the lithium abundance does not decrease
significantly during the MS life time of 4.6 Gyr.

Analysis of helioseismic inversions together with post-model
computations of chemical composition indicates the presence of the
overshooting region and restricts its thickness. It is estimated
to be approximately half of the local pressure scale height (0.5$H_P$)
which corresponds to 3.8\% of the solar radius. Introducing this
extra region does not noticeably deplete lithium during the MS stage. In contrast, at the pMS stage, an
overshooting region with a value of approximately 0.18$H_P$ is
enough to produce the observed lithium depletion.

If we conclude that the dominant lithium burning takes place
during the pMS stage, the dispersion of the lithium abundance in
solar twins is explained by different physical conditions,
primarily during the early stage of evolution before the MS.}


\maketitle
%
%________________________________________________________________

\section{Introduction}

The solar photospheric lithium abundance measured spectroscopically is
160 times less than the solar system's meteoritic abundance (see for example
\citep{Asplund2009}). This discrepancy was already noticed by
\cite{Greenstein1951} and has been widely discussed in the literature (see for
example \citep{Hughes2007} and references therein). The meteoritic
abundance is usually accepted as the initial lithium abundance in
the Sun. Lithium depletion inside a star occurs in thermonuclear
reactions of proton capturing with subsequent nuclear decay into
two $\alpha$-particles: ${\element[][7]{Li}}{\left( {{\mathrm
p},\alpha} \right)}\,\element[][4]{He}$. The solar lithium problem
consists of reproducing the low photospheric abundance as a result
of evolutionary model computations starting from the high initial
value.

If one considers \emph{standard solar modelling}, which includes
only the Main Sequence (MS) stage and one does not consider any
additional mixing beneath the convection zone (CZ), then the
observed lithium abundance in the photosphere cannot be predicted
(see, for example, fig.~1 in \citep{Melendez2010},
\cite{DAntona1984}).  The photospheric
lithium abundance is determined by the temperature and density at
the base of the CZ. This is because matter in the CZ is
efficiently mixed and the abundance of lithium through the whole
CZ is identical and equal to that of the photosphere. The temperature
is the main parameter controlling the nuclear reaction rates, so
the maximum rate of lithium burning is achieved at the base of CZ,
where the temperature is hottest and reaches 2.2~MK in the
present-day Sun (e.g.  model S \citep{Christensen1991,
Christensen1996} defacto accepted as the standard one). But even
at this temperature, the rate of depletion is low, and its
abundance does not decrease significantly during the MS stage.

However, the base of the CZ is remarkably hotter when a radiative
core arises during the early stage of evolution, before the MS.
The maximal temperature in the CZ can reach 3.9 MK  (see, for
example \citep{Piau2002}). As a result, the lithium depletion rate
during the pre-Main Sequence (pMS) is several times higher than
during the MS, athough it still cannot entirely explain the
observed depletion in the frame of standard physics
\citep{Morel1995}. The modelling of the evolution of the lithium
abundance has to take into account the pMS stage where the
depletion is intensive, especially if we observe significant
lithium depletion during the MS stage.

Let us note that non-standard models considering other hypotheses
can lead to changes at the base of the CZ and help to solve the
problem. Several approaches have been proposed to describe extra
mixing: by overshooting \citep{Bohm1963, ahrens92}, by internal
waves \citep{Montalban1994} or by turbulent diffusion
\citep{baglin85, Zhang2012}. This mixing is also influenced by
differential rotation \citep{Zahn1983, Lebreton1987, brun1999} and
magnetic field \citep{McIntyre2007}. Additional lithium depletion
can also be caused by other mechanisms such as stellar wind
\citep{Vauclair1995, Morel1997}, circumstellar disks at the early
stage of evolution \citep{Piau2002} as well as activity during the
early MS lifetime \citep{Turck-Chieze2011}.

Models of the Sun should be consistent with helioseismic data, in
particular with respect to the sound speed profile. Such
discussions on the sound speed profiles in models with
overshooting have been considered in detail by
\cite{Christensen2011, Zhang2012}
(and references therein).  %%%%

In this article, we estimate the rate of lithium depletion of the
present-day Sun using \emph{helioseismic constraints} at the base
of the solar CZ. Usually helioseismic constraints are based on
analysis of the sound speed profile and determination of the depth
of the CZ  \citep{Christensen1991} and allow us to limit the
temperature in the CZ by 2.2 MK \citep{Christensen1996}. Our
analysis method applied to the sound speed \emph{gradient} is much
more sensitive to details of additional mixing and to the gradient
of the hydrogen abundance below the CZ affected by mixing. Our
modelling predicts only small rates of lithium depletion at the
present stage of the solar evolution and this argues in favour of
a dominate role of the pMS stage in lithium evolution.

Sometimes photospheric beryllium abundance is used to test model
computations, but for the Sun, \element{Be} is a much weaker
constraint on the modelling than lithium. Moreover, the
measurement of the beryllium abundance is difficult. A brief
review of this problem can be found in \citep{Andrassy2015}, for
example.

We describe the modelling of the Sun in Sect.~\ref{sect_pMS_MS},
and constrain the region of additional mixing below the CZ in
Sect.~\ref{sect_HelioseismicModel}. Then we analyse the rate of
lithium burning during various stages of the evolution in
Sect.~\ref{sect_pMSmixing}. Our results on solar twins are
discussed in Sect.~\ref{sect_Twins}, before concluding.

%__________________________________________________________________

\section{Standard evolutionary model in CESAM}
\label{sect_pMS_MS}

Evolutionary modelling has been performed using the 1D CESAM2k
code \citep{Morel2008}. Quasi-hydrostatic solar evolution has been
computed until age 4.6 Gyr, including the pMS. The duration of pMS
stage is approximately $ 30$~Myr.

On the basis of the evolutionary computations, we trace the
lithium abundance in CZ. Initial chemical composition of the Sun
is adopted according to \citet{GS1998}, in particular for the
lithium $A(\element[][]{Li})=3.31$. Here, logarithmic abundance is
defined as
$A(\element[][]{Li})=\log(N_{\element[][]{Li}}/N_{\element[][]{H}})+12$
where $N_{\element[][]{Li}}$ and $N_{\element[][]{H}}$ are the
number densities of lithium and hydrogen.

The input physics is relatively common. The temperature gradient
in CZ is calculated using the mixing length theory formalism
\citep{bv1958}. Rates of thermonuclear reactions are computed by
\citet{angulo99}. Microscopic diffusion of a wide set of chemical
isotopes is traced along evolution following methods by
\cite{Michaud1993}. The traced isotopes, besides main species
(hydrogen and helium), are \element[][6]{Li}, \element[][7]{Li},
\element[][9]{Be} and others components of the \element{CNO}
cycle. Equation of state is OPAL2001 \citep{Rogers1996} and
opacity is OPAL1993 \citep{Iglesias1996}. More details on the
standard solar model in CESAM have been given by \cite{Brun1998}.

In the Solar System, lithium abundance is composed of two stable
isotopes \element[][7]{Li} and \element[][6]{Li}. The percentage
of $^7$Li is 92.41\% \citep{Asplund2009}. The main thermonuclear
reactions with lithium are capturing protons with subsequent
nuclear decay: $\element[][7]{Li}{\left( {{\mathrm p},\alpha}
\right)}\,\element[][4]{He}$ and
$\element[][6]{Li}{\left({{\mathrm p},\alpha}
\right)}\,\element[][3]{He}$. The rate of ~\element[][6]{Li}
burning is approximately two orders of magnitude higher than that of
\element[][7]{Li}. Thus, the initial abundance of
\element[][6]{Li} is small and the rate of its burning is high; it
quickly disappeared during the early stages of evolution. Therefore, we
consider only \element[][7]{Li} hereafter.

\begin{figure}
%\resizebox{\hsize}{!}{\includegraphics{fig1.eps}}
\centering
\includegraphics[width=10cm]{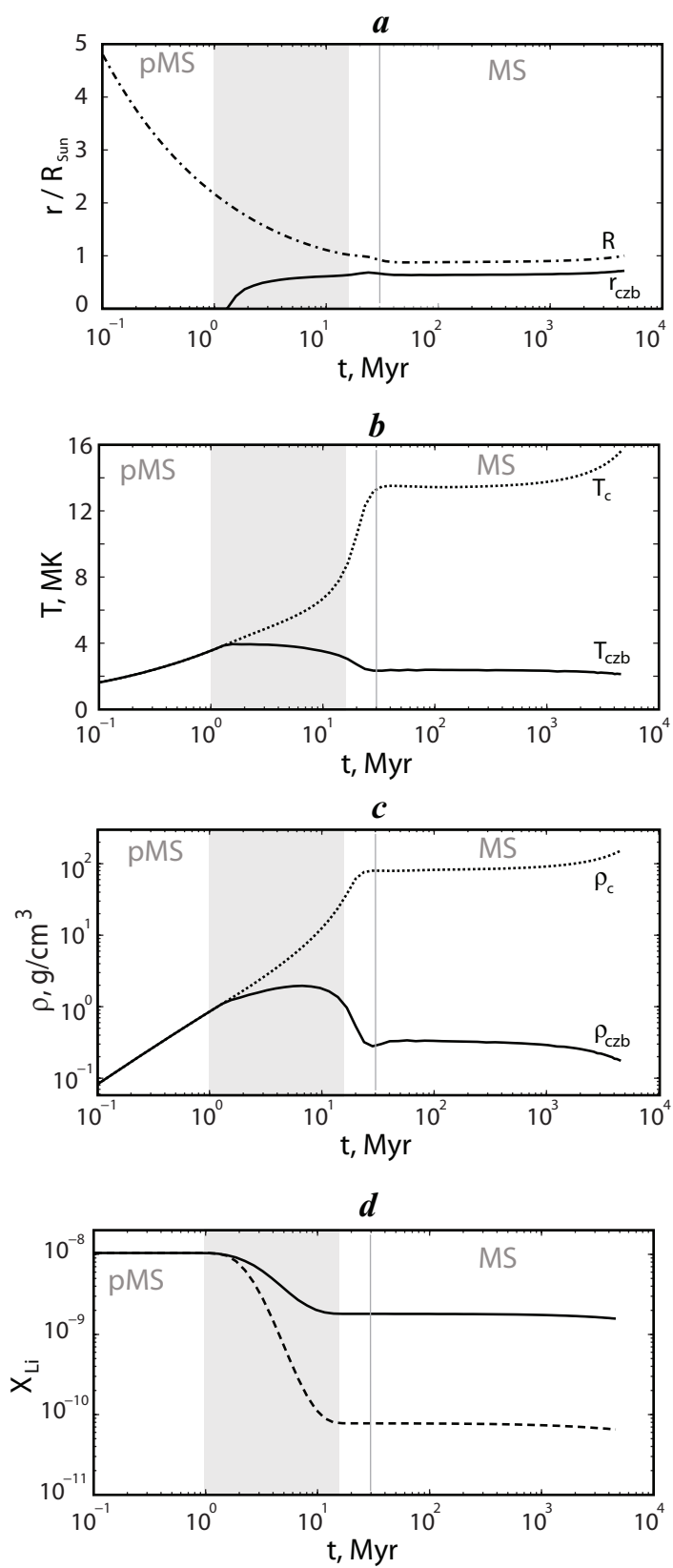}
\caption{Evolution of the Sun during pMS and MS stages. {\bf a} --
solar radius ($R$) and coordinate of the convection zone base
($r_{czb}$); {\bf b} -- temperature at the solar centre ($T_c$)
and at the CZ base ($T_{czb}$); {\bf c} -- density at the solar
centre ($\rho_c$) and at the CZ base ($\rho_{czb}$); {\bf d} --
lithium abundance in CZ in the standard model (solid line) and
with additional mixing (dashed line). Grey vertical lines separate
pMS and MS stages. Grey rectangles highlight time of intensive
lithium burning. } \label{FigMSpMS}
\end{figure}

Two different stages in the solar evolution are considered. The
short initial pMS stage of hydrostatic evolution starts from
collapse of a completely convective star (Fig.~\ref{FigMSpMS}a).
The temperature of the contracting proto-star increases and
achieves approximately 4 MK in the core (Fig.~\ref{FigMSpMS}b),
the matter becomes transparent and convection ceases, as a result
the radiative core appears. It happens at an age of 1.3~Myr. Then,
the radiative-convective star continues to contract and heat up,
the radiative core occupies more space and the convective envelope
becomes shallower \citep{Iben2013}. From view of lithium
evolution, the temperature at the base of CZ is a principal value
and decreases from 3.9 to 2.3 MK during radiative core growth on
pMS. At the same time, behaviour of density at the CZ base is more
complicate and finally reduced to $0.3$~g/cm$^3$
(Fig.~\ref{FigMSpMS}c).

The second stage is a quasi-stationary evolution on the MS which lasts billions of years for the Sun. Stellar
parameters are changing slowly at this stage, particularly the
temperature and position of the base of CZ. On
Fig.~\ref{FigMSpMS}, the grey vertical line conditionally
separates these two stages. It corresponds to the Zero Age Main
Sequence (ZAMS) and characterises change of the source of
luminosity from gravitational contraction to the thermonuclear
reactions \citep{1968psen.book.....C}. At the MS stage, the
convective zone continues very slowly shallowing and the temperature
and density at the CZ base decrease from 2.3 to 2.2 MK and from
0.3 to 0.2 g/cm$^3$ , respectively.

{Our standard computations result in lithium depletion in
the solar CZ by a factor of seven during the whole evolution, while
its main part is depleted during the stage of the pMS (Fig.~\ref{FigMSpMS}d, solid curve).}

%%%%%% Fig. 1 %%%%%%%
%%%%%%%%%%%%%%%%%%%%%

\section{Helioseismic model of slow diffusive mixing
at the stage of MS} \label{sect_HelioseismicModel}

Our main statement is that the rate of lithium burning in the
present solar envelope is too slow to provide the necessary
depletion value. We make this conclusion because the temperature
at the CZ base according to the standard modelling is not high
enough to provide effective lithium depletion, and because
according to our estimation, the thickness of additional mixing
region below CZ could not be extended to reach high-temperature
layers.

The first statement is based on the fact that the temperature at the base
of CZ is fixed by the helioseismic determination
\citep{Christensen1991} of the Schwarzschild point position at the
$r = 0.713{R_\odot }$ and the suggestion that the temperature
profile $T(r)$ in the adiabatic solar CZ is well defined in Model
S \citep{Christensen1996}.

The second statement about additional mixing below the CZ base is
more complex and we outline here the scheme of consideration.

Basic helioseismic information about solar interior is the sound
speed profile $c(r)$ and it is obtained from the eigenfrequencies
with a procedure named helioseismic inversion. The deviation
of the model sound-speed profile from the inverted one is
considered as a measure of quality of the model. But this direct
comparison is adequate in the radiative (i.e. convectively stable)
region below CZ, whereas sound-speed profiles inside adiabatic CZ
are generally very close to one another in different models. Therefore,
models can hardly be distinguished by this criterion.

Instead of analysing the $c(r)$ itself, we use a gradient of sound
speed as a more physically meaningful value. The idea of using the
gradient of sound speed came from early helioseismic work
\citep{Gough1984}, where the author considered the expression
\begin{equation}\label{eq-gradc2}
\frac{1}{g}\frac{d{{c}^{2}}}{dr} \simeq \left( 1-{{\Gamma }_{1}}
\right)
,\end{equation}

\noindent which is well satisfied {inside the adiabatic CZ}.
Here $r$ is a radius-coordinate, $g$ is gravity acceleration and
$\Gamma _1$ is an adiabatic exponent. It is clear from
Eq.~\ref{eq-gradc2} that the gradient of the sound speed is
directly connected with thermodynamic properties via $\Gamma _1$.

Having the sound speed profile, one is able (at least in
principle) to get a gradient of sound speed, but this can be
problematic due to ill-posed numerical differentiation. In our
study, we use results of inversion by \cite{Vorontsov2013} to
acquire an ~`inverted gradient of squared sound speed'.

The first step of our analysis is to obtain information regarding
the gradient of hydrogen abundance $\nabla X$ from the gradient of
sound speed based on a general expression for the gradient of
sound speed via three other gradients. During deduction of the
expression, the differential of $d \ln \rho(\ln P, \ln T, X)$ is
used, and after a number of transformations we acquire
\begin{eqnarray}
\frac{d\ln c^2}{dr} & = & \frac{d\ln \Gamma _1}{dr}\,  -
g\frac{\rho }{P}\frac{\chi _T}{\chi _\rho } \left( \frac{d\ln
T}{d\ln P} - {\left( \frac{\partial \ln T}{\partial \ln P}
\right)}_S \right) + \nonumber \\
 & + & \frac{\nabla X}{X}\frac{\chi _X}{\chi _\rho } -
 \left( \Gamma _1 - 1 \right)\frac{g}{c^2}
 \label{eq-dc2dr}
,\end{eqnarray}

\noindent where $P$ is plasma pressure, $T$ is temperature, $\rho$
is density, $X$ is hydrogen mass fraction and
 $\chi _\rho \equiv {(\partial \ln P / \partial \ln \rho })_{T,X}$,
 $\chi _T \equiv {( \partial \ln P /
\partial \ln T )}_{\rho ,X}$, $\chi _X \equiv {( \partial \ln P /
\partial \ln X )}_{T,\rho }$ are thermodynamic derivatives of the
pressure.

For the region below CZ, we neglect the gradient of the adiabatic
exponent $\Gamma_1$ and suppose that $\Gamma_1$ in the last term
of the right side of Eq.~(\ref{eq-dc2dr}) is adequately prescribed
by the equation of state.

We infer an appropriate gradient of hydrogen from Eq.~\ref{eq-dc2dr},
keeping the gradient of temperature $d\ln T/d\ln P$  as it is in
the model. This inverted ${\nabla X}_\mathrm{inv}$ should provide
a good consistency of model data with helioseismic inversion data.

The second step of our analysis consists of comparison of the
inverted ${\nabla X}_\mathrm{inv}$ with model ${\nabla
X}_\mathrm{mod}$. Then we can conclude whether an additional
mixing occurs beneath the convection zone and we can apply
possible constraints on its value.

Standard evolutionary computations provide model ${\nabla
X}_\mathrm{mod}$ without additional mixing. To study abundance
gradient profile $\nabla X$ in more common cases of mixing, we use
an effective and robust method of post-model computations proposed in
\citep{Baturin2006, Gorshkov2014, Baturin2015}.

The method consists of numerical solution of the evolutionary
equation (\ref{evol-eq}) while the pressure and temperature
profiles are given by some evolutionary sequence and fixed during
integration. In our calculations, model profiles have been adopted
from the Model~S \citep{Christensen1996}.

The evolutionary equation is
\begin{equation} \label{evol-eq}
\rho \frac{{\partial {X_i}}}{{\partial t}} = \, - \nabla  \cdot
\left( {\rho {X_i}\left( {{\alpha _i}\nabla P + {\beta _i}\nabla T
+ {K_D}{\gamma _i}\nabla {X_i}} \right)} \right) + {q_i},
\end{equation}

\noindent where $X_i$ is the mass fraction of  a chemical element,
${\alpha _i},{\beta _i},{\gamma _i}$ are the coefficients of
baro-, thermo-, and concentration diffusion computed following
\citet{Michaud1993} and $K_D$ is a coefficient for describing
mixing. In the standard model without extra-mixing, $K_D = 1$ in
the radiative zone and $K_D>> 1$ in CZ, where mixing is much
faster than diffusion. The last term ${q_i}$  is rate of elements
changing in nuclear reactions. All the values in
Eq.~(\ref{evol-eq}) depend on radius-coordinate $r$ and time $t$.

As a result, a set of ${\nabla X}_\mathrm{mod}$ profiles has been
obtained for various profiles of mixing coefficient $K_D$. Among
this set, we look for a profile of $K_D$ which provides the best
agreement with the inverted profile ${\nabla X}_\mathrm{inv}$.

\begin{figure}
\resizebox{\hsize}{!}{\includegraphics{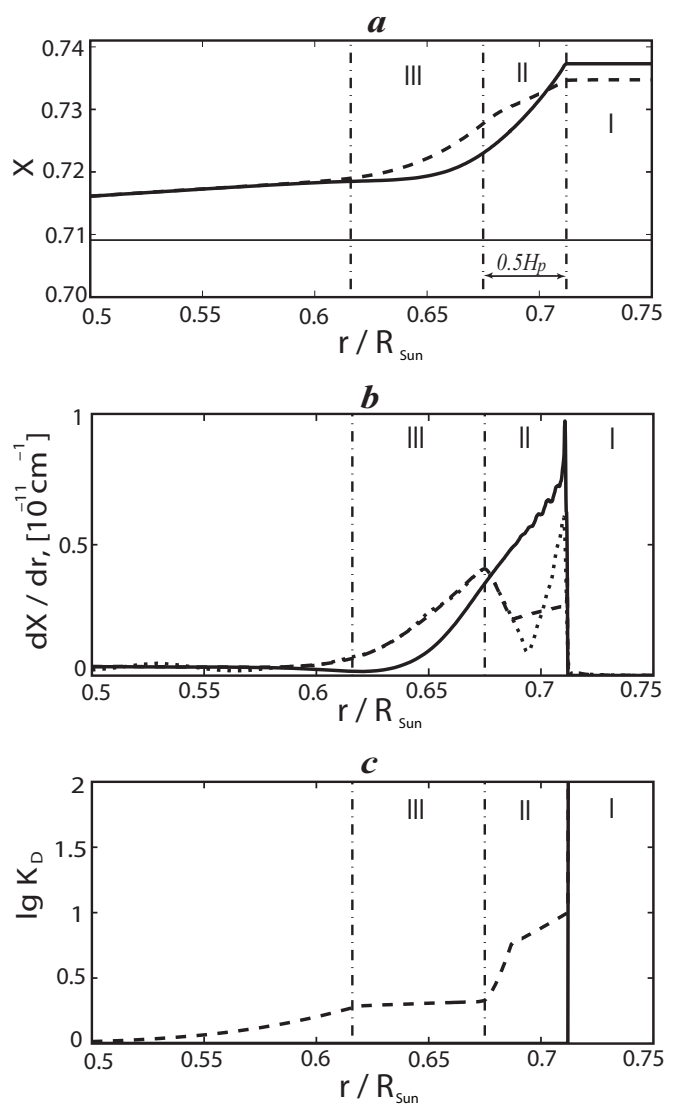}}
\caption{Profiles of hydrogen abundance ({\bf a}), its gradient
({\bf b}) and coefficient $K_D$ ({\bf c}) beneath the convection
zone in the present-day Sun. Dashed and solid lines describe the model
with and without additional mixing, respectively. The thin, solid,
horizontal  line on the plot (a) is hydrogen distribution at the
ZAMS. The pointed line on the plot (b) is inverted
gradient of hydrogen abundance obtained from the observed
sound-speed profile. Vertical dashed-pointed lines separate three
regions with different mixing: I is convection zone, II and III
are zones with additional mixing.} \label{FigXdXK}
\end{figure}

An example of such a comparison is shown by Fig.~\ref{FigXdXK}.
The hydrogen abundance and its gradient resulting from the
matching procedure is presented by dashed curves on
Fig.~\ref{FigXdXK}a,b. The corresponding coefficient ${K_D}$ is on
Fig.~\ref{FigXdXK}c. The convection zone is marked by the roman
numeral I on all panels.

The main results of the matching procedure are the following.
Firstly, below level $r\approx 0.6R_\odot$, agreement between
inverted ${\nabla X}_\mathrm{inv}$ (pointed line) and standard
model ${\nabla X}_\mathrm{mod}$ (solid line) is very good without
assumption of additional mixing. However, in the range from
$0.6R_\odot$ to convection zone base, discrepancy is evident.
There seems to be a necessity for additional mixing in the model.
Secondly, our main conclusion is the possibility to fit model
gradient ${\nabla X}_\mathrm{mod}$ to the inverted ${\nabla
X}_\mathrm{inv}$ in the zone III ($\approx 0.615-0.67R_\odot$) by
supposing slow mixing with $K_D=2$. Thirdly, in the zone II, an
increase of $K_D$ is needed to reach an agreement. However,
further growth of $K_D > 10$ does not provide better agreement in
the vicinity of the CZ base. This high value of $K_D$, however,
means that we reach the fast mixing (overshooting) region. This
region is limited by depth $0.038 R_\odot$, that is 0.5$H_P$,
where $H_P = - (d r/d P)P$ is the local pressure scale height.

%%%%%% Fig. 2 %%%%%%%
%%%%%%%%%%%%%%%%%%%%%

\begin{figure}
\resizebox{\hsize}{!}{\includegraphics{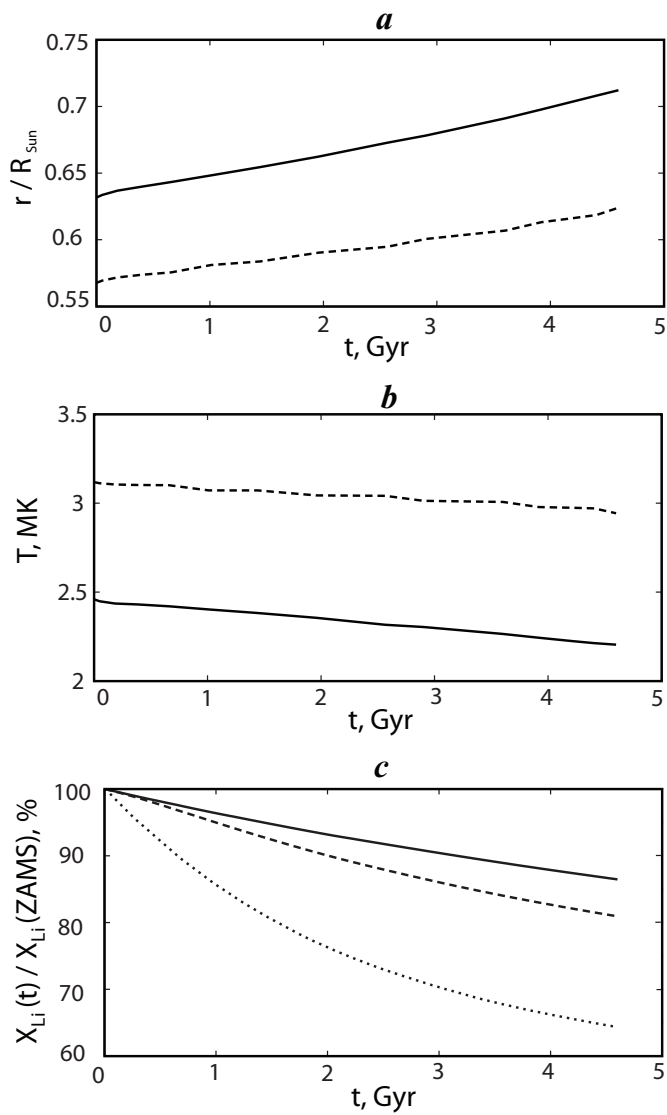}}
\caption{Evolution of the convection zone during the MS stage.
Solid lines indicate classical Model S, dashed ones indicate the
model with additional helioseismic mixing beneath the convection
zone. {\bf a} -- Radius-coordinate of the base of CZ and zone of
the additional mixing, {\bf b} -- temperature at the bases, {\bf
c} -- lithium depletion in both models. The dotted line indicates
lithium depletion in the case of fast mixing.} \label{FigRczbTLi}
\end{figure}

This new increased coefficient of diffusion has been used to
compute the lithium evolution on the Sun. In our model, the
profile $K_D(r,t)$ translates along the radius keeping its shape
with convection zone base evolution. Radius-coordinate of the base
of the convection zone and that of the zone of additional slow
mixing are shown as functions of time in Fig.~\ref{FigRczbTLi}a.
The temperature at the base of the mixing zone is higher than that
at the CZ base by 0.7 MK (Fig.~\ref{FigRczbTLi}b). This leads to
additional lithium burning of approximately 5\%, that is, its
present abundance is 80\% of the initial one
(Fig.~\ref{FigRczbTLi}c, dashed line). Even if we assume mixing in
the poorly known region II to be as fast as in the convection
zone, the envelope lithium abundance decreases insignificantly at
the MS stage. We obtain 65\% of the initial abundance after 4.6
Gyr evolution (pointed line); this is not enough to explain our
observational value.

So, the present helioseismic data do not allow the mixing process
to be deep and fast enough to deplete lithium at the MS stage,
supposing the rate is  constant.

%%%%%% Fig. 3 %%%%%%%
%%%%%%%%%%%%%%%%%%%%%

\section{Fast mixing during the pMS stage}
\label{sect_pMSmixing}

\begin{figure}
\resizebox{\hsize}{!}{\includegraphics{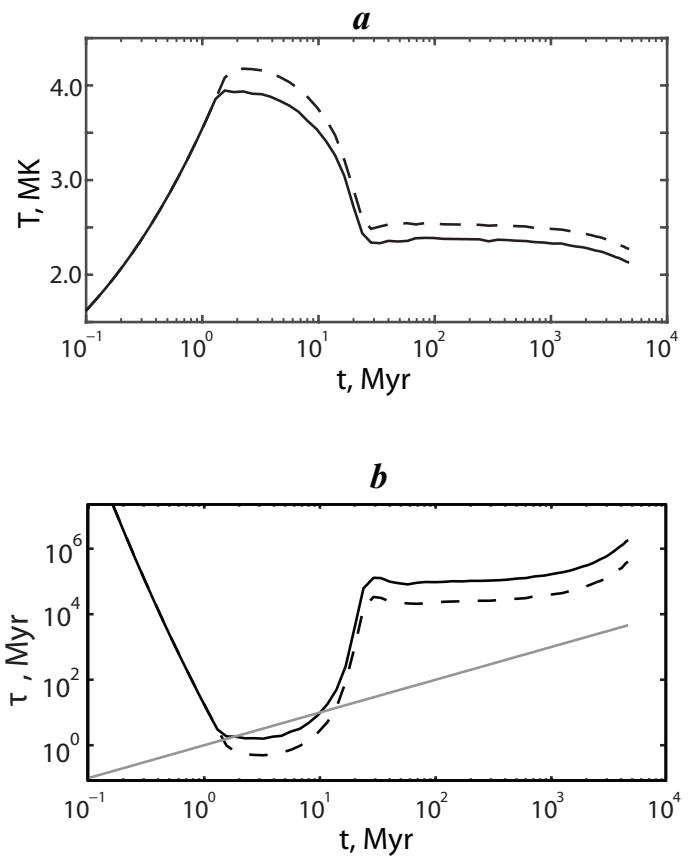}}
\caption{Temperature ({\bf a}) and characteristic times ({\bf b})
for lithium burning in reaction $^7$Li(p, $\alpha$)$^4$He during
the evolution of the Sun. The solid line is obtained for conditions of
the convection zone base, dashed line - for the base of the
overshooting region. The grey straight line corresponds to $\tau$
equal to the current age $t$ of a star. } \label{FigTimes}
\end{figure}

In contrast to the MS stage, an assumption about additional mixing
during the pMS can efficiently explain lithium
depletion in the CZ by a factor of 160 (Fig.~\ref{FigMSpMS}d,
dashed curve). Computations with CESAM2k show that  the depletion
is achieved with a thickness of the extra-mixing region of approximately
$0.18 H_P$. The extra-mixing (convective overshooting) is modelled
assuming a large diffusion coefficient $K_D >> 1$. The thickness
of the overshooting region is constant during the whole evolutionary
computation. The extra-mixing increases lithium
depletion {only slightly during the MS stage}. In contrast, the
rate of lithium burning {during the pMS stage}
is increased significantly because the maximum temperature at the
overshooting region base achieves 4.2 MK, whereas it is 3.9 MK at
the CZ base (Fig.~\ref{FigTimes}a).

Fig.~\ref{FigTimes}b shows the characteristic times $\tau$ of
thermonuclear reaction $\element[][7]{Li}{\left( {{\mathrm
p},\alpha} \right)}\,\element[][4]{He}$:
\[
\tau \equiv
\frac{N(\element[][]{Li})}{{\partial}N(\element[][]{Li})/{\partial
t}},
\]

\noindent where $N(\element[][]{Li})$ is number density of
lithium. The characteristic times are computed for the convection
zone base (solid curve) and for the overshooting region base
(dashed curve). The grey straight line corresponds to
characteristic time $\tau$ equal to the current age $t$ of a star.
When the characteristic time of the reaction is longer than the
current age of a star (i.e. the curve is above the straight line),
the reaction is slow; it does not change the element abundance on
the considered timescale. If the characteristic time is less (the
curve is below the straight line), then the reaction is
significant for the abundance evolution.

We see, firstly, that during the pMS stage, from 1 to 20 Myr, the
overshooting leads to a decrease in characteristic time of lithium
burning by a factor of three. Moreover, it prolongs the time when
the characteristic time of reaction is smaller than the age of a
star. Secondly, during the MS, overshooting value is not important
because characteristic time of burning is much longer than the
current age in both cases.

We may notice that overshooting is not the unique factor which can
lead to efficient lithium burning. Higher opacity, for example,
can play the same role. {The key factor of the lithium
evolution is the maximum temperature at the mixing zone base
during the pMS stage.}

%%%%%% Fig. 4 %%%%%%%
%%%%%%%%%%%%%%%%%%%%%

\section{Solar twins}
\label{sect_Twins}

Solar twins are defined as MS stars with effective temperature,
surface gravity and metallicity close to the solar values, but
independent of age. The mass is assumed to be close to the solar
mass.

We study a set of solar twin stars with measured lithium
abundances. Two different scenarios of lithium depletion in solar
twins during their MS evolution are discussed. The first scenario
considers a low Li depletion rate where A(Li) remains almost
unchanged on the MS: ($|\,\mathrm{d}A(\element{Li})/\mathrm{d}t|
\sim  0.01$ dex/Gyr). We refer to this untraditional scenario as
early depletion evolution course. The second considers a depletion
rate during the MS stage larger than some limit
($|\,\mathrm{d}A(\element{Li})/\mathrm{d}t| \sim 0.2$ dex/Gyr),
and we refer to this as the traditional scenario. Assuming the
traditional scenario, one could expect a noticeable difference in
the lithium abundance between young and old stars.

\begin{figure}
\resizebox{\hsize}{!}{\includegraphics{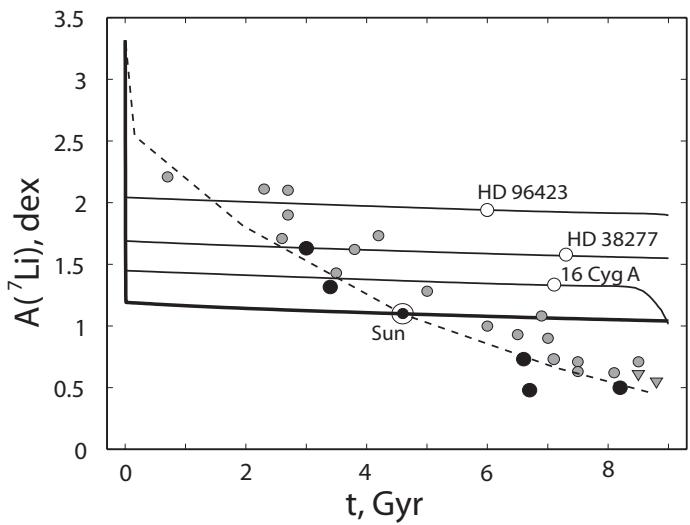}} \caption{
Evolution of lithium abundance in solar twins. Solid lines indicate our
models of early depletion scenario with different overshooting
values (see Table\ref{Table}) leading to significant lithium
depletion during the pMS stage. The dashed line  qualitatively
illustrates the traditional scenario with significant lithium
depletion during the stage of MS. Black circles are solar twins
taken from \citep{Melendez2014}. Grey and unfilled circles and
triangles are twins from \citep{Carlos2016}; triangles are the
upper limits on lithium abundance; three unfilled circles are
twins-exceptions which can be interpreted in the framework of the
early depletion scenario.} \label{FigTwinsMelendezCarlos}
\end{figure}

Li abundances of several solar twins have recently been published.
Observations of this kind could be used to estimate the
MS depletion rate of lithium abundance and could help to
distinguish between the two scenarios. \citet{Melendez2014}
describe five solar twins and these are shown in
Fig.~\ref{FigTwinsMelendezCarlos} (black circles). A similar
analysis is also presented by \citet{Carlos2016}. The authors
demonstrated a strong correlation between A(Li) and the age of 21
selected solar twins. These stars are shown by grey circles and
triangles in the figure. The authors show that the lithium
depletion rate during the MS is high. There are several
evolutionary model predictions in the framework of the traditional
scenario \citep{charbonnel2005, donasci2009, Xiong2009,
denissenkov2010, Andrassy2015}. An example is represented by the
dashed curve.

%%%%%% Fig. 5 %%%%%%%
%%%%%%%%%%%%%%%%%%%%%

We note that the traditional scenario predicts lithium abundance
decreasing rapidly until {the present-day Sun}, in contradiction
with our results, which are based on an analysis of the sound
speed gradient. Comparing the sound speed profiles in the
theoretical models introduced above with the helioseismically
inverted model is not enough to make any conclusions regarding the
depth of the region where additional mixing takes place. In other
words, sound speed profiles may somewhat agree for different
models but the sound speed {gradient} may show specific
peculiarities. We estimate that the rate of lithium depletion is
too small in the Sun and that this therefore favours the early
depletion scenario.

The early depletion scenario does not predict any correlation
between Li and age when starting simulations from a random set of
stars after the pMS. The dispersion of the lithium abundances in
solar twins can be explained through the various physical
conditions during the early pMS stage,  for example, different
overshooting values. Assuming this scenario we can interpret the
solar twins which do not show the strong Li/age correlation, for
example \object{HD 96423}, \object{HD 38277}, and \object{16 Cyg
A} (shown as unfilled circles in
Fig.~\ref{FigTwinsMelendezCarlos}). In Table~1 we list the values
of overshooting that have been adjusted in the stellar evolution
models to match the observed lithium abundances for the given
ages. These values
 do not follow
from any physical model and are merely auxiliary parameters. In
the case of evolution of 16 Cyg A, lithium abundance decreases
rapidly after 8 Gyr because of a hydrogen-fusing shell arising in
this star earlier than in others due to its relatively higher mass.
Thus, the lithium--age strong correlation itself does not reject
the early depletion scenario.

\begin{table}
\caption{\label{Table}Overshooting values used to compute
evolution of the solar twins\tablefootmark{a} }
\begin{center}
\begin{tabular}{lcccc}
\hline\hline
Star & Mass,     & Age,  & $A(\element[][]{Li})$,   & Overshooting,  \\
     & $M_\odot$ & Gyr   & dex                      &  $H_P$   \\
\hline
HD 96423  & 1.03 & 6.0  & 1.93  & 0.117  \\
HD 38277  & 1.01 & 7.3  & 1.58  & 0.140  \\
16 Cyg A  & 1.05 & 7.15 & 1.34 & 0.215   \\
\hline
\end{tabular}
\tablefoot{The masses, ages and observed lithium abundances
$A(\element[][]{Li})$ for HD~96423 and HD~38277 are taken from
\citep{Carlos2016}; for 16~Cyg~A -- from \citep{Ramirez2011}.}
\end{center}
\end{table}

We should also remark that it is hard to explain why a random
sample of solar twins could demonstrate such a {strong}
correlation between lithium abundance and age. The traditional
scenario predicts some correlation between $A(\element{Li})$ and
age, but it is not strong. Statistically it would lead to a strong
correlation only if the dispersion of lithium abundance after the
pMS stage was small meaning that the process of lithium burning
during the pMS stage is strictly identical for all solar twins.
Moreover, the strong correlation implies that the rate of lithium
burning during the MS stage is also well defined, even if stars
have slightly different masses, metallicities and surface gravity
as in the example of \cite{Carlos2016}. As a result, we conclude
that both scenarios do not predict the strong correlation for a
random set of stars.

Considering the different lithium abundance after pMS,
we note that different mixing conditions beneath the CZ
can be explained by, for example, the hydrodynamic
instability below the CZ, which can be affected by rotation as well
as protoplanetary discs \citep{Eggenberger2012}. Observations show
that young stars (up to 500 Myr) can rotate ten times faster than
the Sun \citep{Gallet&Bouvier2013}. The observed dispersion in the
rotational velocities and/or life-time of a disk can explain
different overshooting values in different stars.

Dispersion of pMS lithium abundance can also be caused by
spreading of convective efficiency in the presence of rotation and/or
a magnetic field \citep{Somers&Pinsonneault2014}. Another possibility
is an episodic accretion on young (less than approximately 30 Myr) low-mass stars \citep{Baraffe&Chabrier2010}. After hydrogen burning
has started, stars with the same age and mass may have different
lithium abundance depending on the accretion history. These mechanisms
demonstrate the crucial role of the pMS stage in lithium evolution.

\begin{figure}
\resizebox{\hsize}{!}{\includegraphics{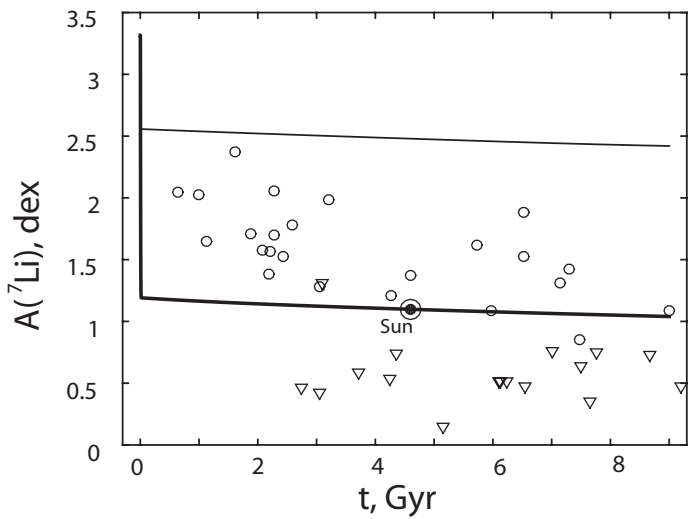}}
\caption{Evolution of lithium abundance in solar twins. Solid
lines show two examples of our computations in framework of early
depletion scenario: thin line shows standard solar model without
overshooting, thick line shows solar model with overshooting
$0.18H_p$. Circles and triangles indicate solar twins considered
by \citep{DelgadoMena2014}, and selected over the ranges of
parameters proposed by \citep{Carlos2016}; triangles are upper
limits on lithium abundance.} \label{FigTwinsDelgado}
\end{figure}

Extensive observations of solar twins (including those re-analysed
by \cite{Carlos2016}) is also provided by \cite{DelgadoMena2014}.
We selected stars with parameters in the range of
\cite{Carlos2016}: $T_\mathrm{eff}=5690-6870$~K, $\log(g) =
4.25-4.50$~dex, $[\element{Fe}/\element{H}]=-0.11-(+0.11)$, and
also $M=(0.94-1.07)M_\sun$. The selected stars are shown in
Fig.~\ref{FigTwinsDelgado}, presenting different lithium
abundances in both young and old stars. Moreover, there is no
significant correlation between A(Li) and age. In the context of
early depletion scenario they can be explained by various physical
conditions during the early stage of evolution, for example, by
different overshooting values as shown in
Fig.~\ref{FigTwinsDelgado}, or different A(Li) in the clouds from
which stars are born.

Discussions about determining ages of stars of \cite{Carlos2016}
and \cite{DelgadoMena2014} are very interesting but this is
outside the scope of this work.

%%%%%% Fig. 6 %%%%%%%
%%%%%%%%%%%%%%%%%%%%%

\section{Conclusions}
\label{sect_Conclusions}

Considering the general evolution of solar-type stars, we make
conclusions about the prevailing role of the early pMS stage on the
present-day lithium abundance in the convective envelope.

In the stellar surface envelope, the lithium abundance decreases by a
factor of seven during the whole evolution in models without
extra mixing while the lithium is mainly depleted during the pMS
stage.

When we assume the mixing model under the CZ in the present-day
Sun restricted by helioseismic analysis, the lithium depletion
increases only by 5\%  during the MS stage.

Low observed lithium abundance on the Sun can be explained by the
existence of regions with additional mixing during the early stage
of evolution, before the MS stage. The thickness of extra-mixing
regions is estimated as 0.18$H_P$.

Dispersion of lithium abundance in solar twins could be explained
by variation of physical conditions during the pMS stage whilst keeping
lithium abundance almost stable during the MS stage.

\begin{acknowledgements}
      Part of this work was supported by a research project
      ``Mod\'elisation Cesam2k''  of Observatoire de
      la C\^ote d'Azur.
\end{acknowledgements}

\bibliographystyle{aa} % style aa.bst
\bibliography{Thevenin_bibliogr} % your references Yourfile.bib

\listofobjects

\end{document}